\documentclass[conference]{IEEEtran}
\IEEEoverridecommandlockouts

\usepackage{bbm}
\usepackage{graphicx}

\usepackage{xcolor}
\usepackage{float}   
\usepackage[normalem]{ulem}
\usepackage{dsfont}
\usepackage{textcomp}
\usepackage{stfloats}
\usepackage{url}
\usepackage{verbatim}
\usepackage{algorithmic}
\usepackage{array}
\usepackage{booktabs}
\usepackage{pifont}
\usepackage{comment}
\usepackage{algorithm}
\usepackage{algorithmic}
\usepackage[acronym]{glossaries}
\newacronym{3gpp}{3GPP}{3rd Generation Partnership Project}
\newacronym{aoi}{AoI}{Age of Information}
\newacronym{ack}{ACK}{ACKnowledgment}
\newacronym{agv}{AGV}{Automated Guided Vehicle}
\newacronym{bs}{BS}{Base Station}
\newacronym{cdmp}{CDMP}{Constrained Markov decision process}
\newacronym{cowu}{CoWu}{Content-based Wake-up}
\newacronym{coud}{CoUD}{Cost of Update Delay}
\newacronym{csma}{CSMA}{Carrier Sense Multiple Access}
\newacronym{dl}{DL}{downlink}
\newacronym{embb}{eMBB}{enhanced Mobile Broadband}
\newacronym{goe}{GoE}{Grade of Effectiveness}

\newacronym{iiot}{IIoT}{Industrial Internet of Things}
\newacronym{iot}{IoT}{Internet of Things}
\newacronym{id}{ID}{IDentity}
\newacronym{idwu}{IDWu}{Identity-based Wake-up}
\newacronym{k-qaoi}{$k$-QAoI}{top-$k$ QAoI}
\newacronym{kpi}{KPI}{Key Performance Indicator}

\newacronym{mmtc}{mMTC}{Massive Machine-type Communications}

\newacronym{pdf}{PDF}{probability density function}
\newacronym{ppp}{PPP}{Poisson Point Process }
\newacronym{qaoi}{QAoI}{Query AoI}
\newacronym{pmf}{PMF}{probability mass function}

\newacronym{rr}{RR}{round-robin scheduling}

\newacronym[shortplural=SAs-Push, longplural=Sensing Agents for Push-based communication]{sa-push}{SA-Push}{Sensing Agent for Push-based communication}
\newacronym[shortplural=SAs-Pull, longplural=Sensing Agents for Pull-based communication]{sa-pull}{SA-Pull}{Sensing Agent for Pull-based communication}
\newacronym{ucwu}{UCWu}{UniCast Wake-up}
\newacronym{ul}{UL}{uplink}
\newacronym{urllc}{URLLC}{Ultra-Reliable and Low Latency Communications}
\newacronym{wus}{WuS}{wake-up signal}

\newacronym{wsn}{WSNs}{Wireless Sensor Networks}
\newacronym{mac}{MAC}{Medium Access Control}
\usepackage{amsmath,amsfonts, amssymb}
\usepackage{bm}
\usepackage{graphicx}

\usepackage[font=small]{caption}
\usepackage[font=footnotesize]{subcaption}
\usepackage{tikz}
\usetikzlibrary{plotmarks,patterns,decorations.pathreplacing,backgrounds,calc,arrows,arrows.meta,spy,matrix}
\usepackage{tikzscale}

\usepackage{pgfplots}
\usepgfplotslibrary{colorbrewer, groupplots, patchplots}
\pgfplotsset{
    compat=newest,
    legend style={font=\footnotesize, fill opacity=0.7,  draw opacity=1, text opacity=1, draw=white!15!black, legend cell align=left, align=left}, 
    width=0.8\columnwidth, 
    scale only axis,
    height=4cm,
    yminorticks=false,
    xminorticks=false,
    label style={font=\small},
    title style={font=\small},
    tick align=outside,
    tick pos=left,
    tick style={color=black},
    tick label style={font=\footnotesize},
    grid style={line width=.1pt, draw=gray!20},
    major grid style={line width=.1pt,draw=gray!20},
    plot coordinates/math parser=false % not sure if useful
}
\newlength\figureheight
\newlength\figurewidth

\usepackage{multirow}
\usepackage{tablefootnote}
\usepackage{booktabs}
\usepackage{tabularx}

\def\BibTeX{{\rm B\kern-.05em{\sc i\kern-.025em b}\kern-.08em T\kern-.1667em\lower.7ex\hbox{E}\kern-.125emX}}

\newcommand{\mb}[1]{\mathbf{#1}}    % mathbold abbreviation

\newcommand{\argmin}{\mathop{\rm arg~min}\limits}
\newcommand{\probP}{\text{I\kern-0.15em P}}
\newcommand{\T}{^{\mathsf{T}}}     % transpose
\definecolor{amaranth}{rgb}{0.9, 0.17, 0.31}

\begin{document}

% \title{Analysis of Content-based Wake-up for Pull and Push Coexistence in IoT Networks}
\title{Coexistence of Push Wireless Access with Pull Communication for Content-based Wake-up Radios}

\author{\IEEEauthorblockN{%    
    Junya Shiraishi\IEEEauthorrefmark{1},~Sara Cavallero\IEEEauthorrefmark{2},~Shashi Raj Pandey\IEEEauthorrefmark{1},~Fabio Saggese\IEEEauthorrefmark{1}, and~Petar Popovski\IEEEauthorrefmark{1} \medskip
    }
    \IEEEauthorblockA{%
    \IEEEauthorrefmark{1}Department of Electronic Systems, Aalborg University, Denmark. 
    \IEEEauthorrefmark{2}WiLab/CNIT and University of Bologna, Italy. \\
    \IEEEauthorrefmark{1}\{jush, srp, fasa, petarp\}@es.aau.dk,
    \IEEEauthorrefmark{2}s.cavallero@unibo.it}
    \thanks{This work was partly supported by the Villum Investigator Grant ``WATER" from the Velux Foundation, Denmark, partly by the Horizon Europe SNS ``6G-XCEL" project with Grant 101139194, and partly by the  Horizon Europe SNS ``6G-GOALS'' project with grant 101139232.}
    }
%\date{January 2024}

\maketitle

\begin{abstract}
This paper considers energy-efficient connectivity for Internet of Things (IoT) devices in a coexistence scenario between two distinctive communication models: pull- and push-based. In pull-based, the base station (BS) decides when to retrieve a specific type of data from the IoT devices, while in push-based, the IoT device decides when and which data to transmit. To this end, this paper advocates introducing the content-based wake-up (CoWu), which enables the BS to remotely activate only a subset of pull-based nodes equipped with wake-up receivers, observing the relevant data. In this setup, a BS pulls data with CoWu at a specific time instance to fulfill its tasks while collecting data from the nodes operating with a push-based communication model. The resource allocation plays an important role: longer data collection duration for pull-based nodes can lead to high retrieval accuracy while decreasing the probability of data transmission success for push-based nodes, and vice versa. Numerical results show that CoWu can manage communication requirements for both pull-based and push-based nodes while realizing the high energy efficiency (up to 38$\%$) of IoT devices, compared to the baseline scheduling method.
\end{abstract}

\section{Introduction}
\label{sec:intro}
In the future generation of wireless communication, energy and spectrum efficiencies are considered among the fundamental \glspl{kpi} for \gls{iot} networks~\cite{Chettri2020:iot-survey}. It is then imperative to design communication protocols that support \emph{sustainable connectivity} between different \gls{iot} devices, serving different communication classes, service requirements, and tasks. 
To this aim, recent research brings focus on \emph{goal-oriented communications} as a promising approach, in which communication is designed as a means to achieve certain goals of the communicating parties~\cite{Goldreich2012:goal-theory, gunduz2022beyond}. In \gls{iot} networks, this principle can be translated into transmission policy that focuses on information relevant for receiver's current task, which reduces unnecessary data transmission and thereby increases the overall energy efficiency. Accordingly, the data receiver, e.g., the \gls{bs}, can strategically decide when and which devices to query for relevant data. This communication regime, named \emph{pull-based} communication, has recently gained attention~\cite{Chiariotti2022:qAoI, shiraishi2022query}. It differs from the classical \emph{push-based} communication, in which the devices generating data decide autonomously when and what to transmit. 

The technology enabler for low-energy pull-based \gls{iot} communication is the \emph{wake-up radio}, an ultra low-power wake-up receiver installed into the \gls{iot} devices that remains active waiting for specific \glspl{wus} while the higher energy-demanding main radio is turned off~\cite{piyare2017ultra}. Therefore, this allows devices to operate in a demand-driven manner, i.e., upon a received \gls{wus} from the \gls{bs}, the device activates its main radio and transmits its observations, reducing the energy consumption during idle periods~\cite{IEICE-yomo}. 
The design of the \glspl{wus} is currently studied in the \gls{3gpp} standardization body~\cite{3GPP_standard,wagner2023low,hoglund20243gpp}, which considers the \gls{idwu} as a viable solution. In \gls{idwu}, the \gls{bs} transmits a wake-up signal embedding the unique \gls{id} of a device~\cite{piyare2017ultra}. However, employing \gls{idwu} cannot avoid the wasteful wake-up of nodes which might transmit data irrelevant to the current task, deteriorating the energy efficiency of the \gls{iot} devices. 
To address this problem, we have proposed \gls{cowu}~\cite{TGCN_Content}. In \gls{cowu}, the information about the data needed by the \gls{bs} is embedded into the \gls{wus}, letting only the nodes having suitable readings turn on their main radio and transmit~\cite{TGCN_Content}. However, in practical \gls{iot} scenarios, nodes operating with pull-based and push-based communication models might coexist~\cite{Sara_Pull_and_push}. For example, in a digital twin setting, the \gls{bs} might constantly collect status updates from the push-based nodes to maintain the accuracy of the digital twin while it directly requests critical data in a pull-based manner to control the elements in the physical twin, such as \gls{agv} or mobile actuator. In these cases, designing the \gls{mac} layer frame in which push-based and pull-based communications coexist is also fundamental. In this regard, the authors~\cite{talli2024push} described an analytical formulation for choosing the two communication modes from the \gls{bs} perspective. Similarly, in a recent work ~\cite{Sara_Pull_and_push}, \gls{mac} layer access problem for \gls{iot} data collection in the coexistence of \gls{idwu} pull-based and push-based communication was addressed. 

Related works thereof indicate a general problem within the coexistence of pull-push communication. Activating pull-based nodes to transmit often for goal-oriented objectives at the \gls{bs} eventually deteriorates the performance of push-based nodes due to fewer transmission opportunities within the shared communication resources. This motivates the objective of this work: \emph{how to design an efficient \gls{mac} protocol for pull/push coexistence in \gls{iot} data collection scenarios to realize sustainable connectivity}. To the best of our knowledge, no previous work has considered applying \gls{cowu} for the pull/push communication coexistence scenario.

This paper investigates the performance of \gls{cowu} in particular, proposing a mechanism to allow the coexistence of the two traffics. In this regard, our contributions are two-fold: first, we characterize the basic trade-off between the accuracy of the retrieved data of pull-based communication and the probability of successful access for push-based communication through theoretical analysis; second, we evaluate the gain of applying \gls{cowu} against a scheduling method, showing its robustness against the traffic of push nodes.

\section{System Model}
\label{sec:system_model}
\begin{figure}[t]
    \centering
    \begin{subfigure}{0.99\columnwidth}
        \centering
        \includegraphics[width=\textwidth]{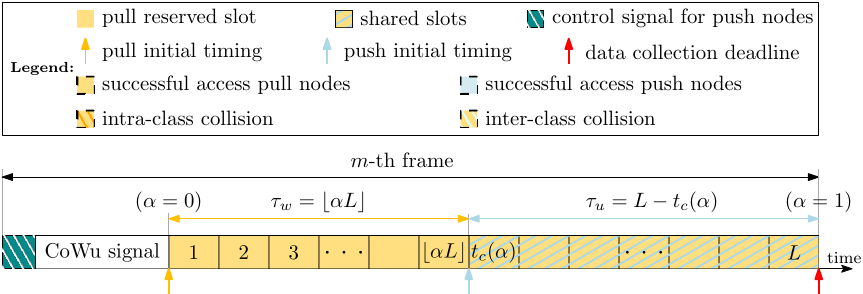}
        \caption{\gls{mac} frame structure.}
        \label{fig:cowu:frame}
    \end{subfigure}
    \begin{subfigure}{0.99\columnwidth}
        \centering
        \includegraphics[width=\textwidth]{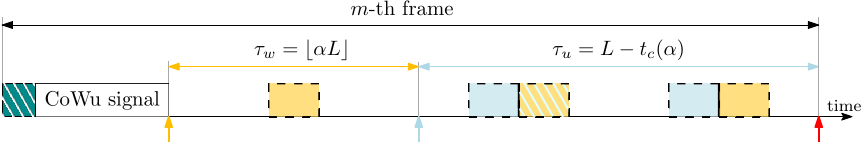}
        \caption{Shorter $\tau_{w}$ for \glspl{sa-pull}. Inter-collision happens between \glspl{sa-pull} and \glspl{sa-push}.}
        \label{fig:cowu:shorter}
    \end{subfigure} 
    \begin{subfigure}{0.99\columnwidth}
        \centering
        \includegraphics[width=\textwidth]{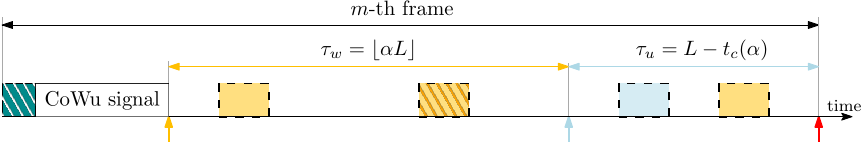}
        \caption{Longer $\tau_{w}$ for \glspl{sa-pull}. Intra-collision happens between \glspl{sa-pull}.}
        \label{fig:cowu:longer}
    \end{subfigure}       
    \caption{An example of operation when applying \gls{cowu} for pull and push coexistence scenario.}%, where 3 \glspl{sa-pull} are woken up by \gls{cowu} signaling.}
    \label{fig:cowu_coexistence}
    \vspace{-4mm}
\end{figure}

We consider a scenario of \gls{iot} data collection in the coexistence of pull and push-based communication, where the \gls{bs} collects data from both \glspl{sa-pull} and \glspl{sa-push}. The number of \glspl{sa-pull} and \glspl{sa-push} are denoted as $N_{w}$, and $N_{u}$, respectively. All the devices access the channel following a shared \gls{mac} protocol, which divides the time horizon into frames of equal length. The proposed structure of the frame is illustrated in Fig.~\ref{fig:cowu:frame}. Each frame starts with \gls{dl} transmissions: the \gls{bs} sends control information regarding the structure of the frame to the \glspl{sa-push}; then, it transmits a \gls{cowu} signal to retrieve specific data from the \glspl{sa-pull}. Details on the content of the \gls{dl} signaling are given in Sec.~\ref{sec:model:pull} and~\ref{sec:model:push}. The remaining part of the frame is divided into $L$ time slots used for \gls{ul} transmission, each of duration $T_s$ [s]. Furthermore, the \gls{ul} portion of the frame is divided into two parts: $\tau_w$ slots reserved for pull and $\tau_u$ shared slots, such that $\tau_w + \tau_u = L$. 
This indicates only \glspl{sa-pull} can transmit within the firsts $\tau_w$ slots, while both \glspl{sa-pull} and \glspl{sa-push} can attempt the access in the subsequent $\tau_u$ slots.
The \gls{bs} controls the ratio between the slots reserved for pull and shared by tuning the parameter $\alpha\in[0,1]$. Specifically, in each frame, the set of time slots $\{1, \dots, \lfloor \alpha L \rfloor\}$ having length $\tau_w = \lfloor \alpha L \rfloor$, is reserved for \glspl{sa-pull} access, while the set of slots $\{t_c(\alpha), \dots, L\}$, having length $\tau_u = L - t_c(\alpha)$, is shared among all the devices,  where $t_{c}(\alpha) = \min(L, \lfloor{\alpha}L{\rfloor}+1)$. Remark that for $\alpha = 0$, no pull reserved slots are present, i.e., $\tau_w = 0$ and $\tau_u = L$ while for $\alpha= 1$, no shared slots are present, i.e., $\tau_w = L$ and $\tau_u = 0$. 
%$(\tau_w, \tau_u) = (0, L)$ 

Regardless of the types of agents, we assume a collision channel for \gls{ul} communication and an error-free channel for the \gls{dl} direction. To simplify the analysis, we assume that all devices employ a $p$-persistent \gls{csma}: when a device attempts to access the channel, it first conducts carrier sensing at the beginning of a slot. If the channel is idle, the node attempts transmission of its packet with transmission probability $p$. Each transmission attempt occupies a time slot. An error-free \gls{ack} is received from the \gls{bs} immediately after each successful transmission, as in \cite{Sara_Pull_and_push}. The nodes detecting a packet loss by the absence of the \gls{ack} re-transmit their packets following $p$-persistent \gls{csma}. Here, we assume that all nodes, including the \gls{bs}, are within each other's communication, wake-up, and carrier-sensing range. Accordingly, two types of collisions may happen: intra-class and inter-class collisions. The former happens among the data transmission operating in the same communication mode, e.g., pull-based with pull-based communication, whereas the latter happens among the nodes operating in different modes. While intra-class collisions are independent of the frame structure, if the \gls{bs} reduces $\tau_{w}$ (see Fig.~\ref{fig:cowu:shorter}), the data transmission from the push-based nodes would succeed with high probability while there is a high risk that the data transmission from the pull-based nodes will fail due to inter-class collisions. On the other hand, if the \gls{bs} increases $\tau_{w}$ (see Fig.~\ref{fig:cowu:longer}), the transmission of the active pull-based nodes is highly likely to succeed, but the performance of push-based nodes deteriorates due to the fewer transmission opportunities.

\subsection{Model of pull-based communication}
\label{sec:model:pull}
Each \gls{sa-pull}, $n\in\{1,\dots, N_w\}$, is equipped with a wake-up receiver, and monitors an independent and identically distributed physical random process $V_{n}[m]\in [V_{\mathrm{min}}, V_{\mathrm{max}}]$, whose value is assumed to be approximately constant during the frame $m$, and distributed according to a \gls{pdf} $g_{V}(v; m)$. When the \gls{bs} receives an external query via cloud -- for example, from a mobile actuator -- it retrieves data from the \glspl{sa-pull} through the transmission of wake-up signals. The queries arrive at the \gls{bs} in a sporadic manner and contain a specific deadline $T$ for the data collection. For simplicity of analysis, we assume that the deadline is set to the end of the frame. In this paper, we focus on the system-level behavior in a single $m$-th frame, assuming the \gls{bs} received a query request in the previous $(m-1)$-th frame.
% receives data from $N_{w}$ \gls{iot} devices with a wake-up receiver deployed over the sensing field by wake-up signaling. 
%When the \gls{bs} receives a query requests at a $(m-1)$-th frame, it transmits a \gls{cowu} signal at the beginning of $m$-th frame to collect data from the \gls{sa-pull}.

In general, the query request could be for a $d-$ dimensional content $\mathcal{X}~\in~\mathbb{R}^{d}$, where $d \geq 1$. Without loss of generality, in this paper, we focus on the case with $d = 1$ and assume the query asks the data from observations within the range $V_{L} \le V_n[m] \le V_{U}$, $\forall n \in\{1, \dots, N_w\}$. In other words, the \gls{bs} has to conduct the so-called range-query employing a \gls{cowu} signal which embeds the information of target range threshold\footnote{This can be realized, e.g., by encoding the lower and higher values of a range of interests into the length of wake-up signal~\cite{shiraishi2022query, TGCN_Content}. In this work, without loss of generality, we ignore the impact of the variable length of the \gls{cowu} and restrict our analysis to a given number $L$ of \gls{ul} available slots.} $\mb{V}_\mathrm{th}= [V_L, V_U]$. When a wake-up receiver receives the \gls{cowu} signals, the \gls{sa-pull} samples its observed physical process and checks whether its observation is within the range threshold $\mb{V}_\mathrm{th}$~\cite{shiraishi2022query, TGCN_Content}. If so, it activates the main radio and attempts to access the channel transmitting a single packet. % Thus, \gls{cowu} enables the \gls{bs} to activate \glspl{sa-pull}, which are highly likely to belong to \JS{range} set at the current frame and can suppress the data transmission from the others. 
Consequently, if a \gls{sa-pull} receives an \gls{ack} from the \gls{bs} signaling successful access, it turns off its main radio interface. Therein, the energy consumption of the \glspl{sa-pull} is evaluated according to the following assumptions. When no communication is requested, each \gls{sa-pull} consumes power only due to the wake-up receiver, which is negligible. On the other hand, the \glspl{sa-pull} that are activated by the wake-up signal are either in the transmit or in the receiving state; we denote the power consumption in the two states by $\xi_{T}$ and $\xi_{R}$, respectively. Accordingly, the energy consumption of a \gls{sa-pull} is directly related to the amount of time used attempting the access or listening to the channel, as detailed in Sec.~\ref{sec:analysis:coexistence}.

In this setting, we can define the \emph{accuracy of pull retrieved data}, namely $\gamma_w$. Denoting by $\mathcal{T}$ the subset of \glspl{sa-pull} observing the process within the target range threshold $\mb{V}_\mathrm{th}$, and by $\mathcal{S}$ the subset of \glspl{sa-pull} successfully transmit the data by the end of the frame, the accuracy is
\begin{equation}
    \gamma_w = \mathrm{Pr}(\mathcal{T}=\mathcal{S}).
    \label{eq:range_set}
\end{equation}
The accuracy becomes 1 if the \gls{bs} collects data from all \glspl{sa-pull} through wake-up signaling, which, however, may sacrifice the communication from the \glspl{sa-push}, and also increases the total energy consumption of the \glspl{sa-pull}.

\subsection{Model of push-based communication}
\label{sec:model:push}
\Glspl{sa-push} are intermittently active, taking measurements of the environment according to a \gls{ppp} and generating packets to transmit toward the \gls{bs} with an average arrival rate $\lambda$ [packet/slot]. Arguably, one can consider the \gls{bs} queries ``all contents" of the \glspl{sa-push}, while only specific content of \glspl{sa-pull} through the \gls{cowu} signal; thus, reaching a unified understanding of how the \gls{bs} can collect data. At the beginning of the frame, the \gls{bs} notifies the time periods $[t_{c}(\alpha), L]$, during which \glspl{sa-push} can transmit its packet inserting this information into the control signal for push nodes. For simplicity of analysis, we assume that in each frame, \glspl{sa-push} store only the latest packet in the queue and transmit it in the next frame, if any. Thus, the probability that a \gls{sa-push} attempts data transmission in the frame $m$ is $p_{\lambda} = 1 - e^{-\lambda L}$, which is the probability that a node generates at least one packet during the $m-1$-th frame. 

%\fs{For the \glspl{sa-push}, we ignore the energy consumption metric, considering that the push-based communication acts independently, and the proposed coexistence design cannot influence the policy of attempting the access transmission.}
%\JS{For the \glspl{sa-push}, we ignore the energy consumption metric, assuming that all \glspl{sa-push} observes mission critical information and their main radio are kept active in order to reduce the communication latency. In other word, we defer the consideration of any energy management techniques including duty-cycling and wake-up radio for \glspl{sa-push}. The investigation of data collection policy considering both energy consumption of \glspl{sa-pull} and \glspl{sa-push} are kept for future work.}
Here, as in the scope of this paper, we look at the access mechanism for reducing the energy consumption for \gls{sa-pull} while satisfying the requirements for both \glspl{sa-pull} and \glspl{sa-push}. Thus, we only focus on the \emph{probability of successful access} for \glspl{sa-push} - or just \emph{push success probability}, - $\gamma_u$, as the main metric for the push-based communication thereof, ignoring the energy consumption for \gls{sa-push} and keep it for future works.  
%\fs{Denoting as $u \le N_u$ the number of \glspl{sa-push} generating a packet in the $m-1$-th frame, and $u' \le u$ the number of packets correctly received by the \gls{bs} at the end of the $m$-th frame, the push success probability is}
%\fs{Denoting as $u \le N_u$ the number of \glspl{sa-push} generating a packet in the $m-1$-th frame, and $u' \le u$ the number of packets correctly received by the \gls{bs} at the end of the $m$-th frame, the push success probability is}
%\begin{equation}
%    \fs{\gamma_u = \Pr\left( u = u' \right).}
%\end{equation}
Therein, $\gamma_u$ is defined as the probability that data transmission of \gls{sa-push} is succeeded in the $m$-th frame. %under the condition that a \gls{sa-push} generates a packet in the $m-1$-th frame.} %Specifically, let us denote the binary variable representing whether a \gls{sa-push} has a packet at the beginning of $m$-th frame as $a$, where $a= 1$ means the nodes has a packet, and $a = 0$ is not and the one representing whether data transmission }
%\begin{equation}
%    \fs{\gamma_u = \Pr\left(h= 1 |a = 1 \right).}
%\end{equation}

\section{Analysis}
\label{sec:analysis}

In this section, we derive the accuracy of the retrieved range set and the total energy consumption for \gls{sa-pull}, and the probability of successful access for \gls{sa-push}.

\subsection{Analysis of the distribution of successful transmissions}
\label{sec:analysis:successful-nodes}
To characterize the system level performance, it is required to obtain the distribution of the number of successful data transmissions at the end of the portion of the frame reserved for pull, where only intra-class collision can happen, and the end of the portion of shared resources, where both intra- and inter-class collisions can happen. In other words, we are interested in finding the number of successful transmission attempts happened from slot 1 to slot $\lfloor\alpha L\rfloor$, and the successful attempts from slot $t_c(\alpha)$ and slot $L$.

To analyze the aforementioned distribution in a general manner, we construct a Markov chain in which each state $s$ represents the number of packets available for transmission. Considering $J$ packets to be transmitted at any time instance within the frame, the state space is defined as $\{J, J-1, \ldots, 1, 0\}$. The transition from state $s$ to state $s-1$ happens if only one successful transmission exists for a node -- with probability $p$, -- while all the other nodes do not attempt transmission -- each with probability $1-p$.  Accordingly, the transition probability $p_{s, s-1}$ for $s \in\{1, \dots, J\}$ is
\begin{equation}
    p_{s, s-1}=s\,  p\,  (1-p)^{s-1}.
    \label{eq:succ_pull}
\end{equation}
On the contrary, the probability of remaining in the same state can be described as $p_{s, s} = 1 - p_{s, s-1}$,  $\forall s\in\{1,\dots, J\}$. %The transition probability from $s$ to $s+1$ \fs{is null \forest{as the number of packets available for transmission is defined at the beginning of the frame and only decreases with each successful transmission within that frame. That means any packet generated in a frame can be sent only in the next frame.} because any packet generated in a frame can be sent only in the next frame}. 
Finally, state $s = 0$ is an absorbing state representing the event all $J$ packets have been successfully transmitted. Note that the other transition probabilities are 0, according to  Sec.~\ref{sec:system_model}.

We now define the state vector $\mathbf{\Phi}(t)= [\phi_J(t), \dots, \phi_s(t) \dots, \phi_{0}(t)]\T\in[0,1]^{(J +1)}$ whose elements represent the probability of every state $s$ after $t$ time slots from a specific time instance, which, in our case, is either the beginning of the pull reserved slots (slot $1$) or the shared slots (slot $t_c(\alpha)$). %whose $s$-th entry corresponding to state $s$ is denoted as $\mathbf{\phi}_{s}(t)$ 
The dynamic of the state vector through time results in: 
\begin{equation}
    \mathbf{\Phi}(t+1) = \mathbf{\Phi}(t) \, \mathbf{R},
    \label{eq:power_method}
\end{equation} 
where $\mathbf{\Phi}(0)$ is the initial state distribution, defined as $\mathbf{\Phi}(0) = [1, 0, 0, \ldots, 0]\T$, and $\mathbf{R} \in \mathbb{R}^{(J+1) \times (J+1)}$ is the transition matrix containing the transition probabilities. Using eq.~\eqref{eq:power_method}, the probability that $j$ out of the $J$ available packets at $t=0$, succeed after $\zeta$ time slots is expressed as follows:
\begin{equation}
P_{s}(j | J,  \zeta)=\phi_{J-j} (\zeta).\label{eq:n_s_D}
\end{equation}

\subsection{Analysis of pull/push coexistence metrics employing \gls{cowu}}
\label{sec:analysis:coexistence}
Let $P_{w}(\mb{V}_\mathrm{th})$ denote the probability of \glspl{sa-pull} waking up given the \gls{cowu} threshold $\mb{V}_\mathrm{th}$. Given $g_{V}(v;m)$ the \gls{pdf} of the observed process at the $m$-th frame, we have
\begin{equation}
P_{w}(\mb{V}_{\mathrm{th}})=\int_{V_{L}}^{V_{U}}g_{V}(v; m)dv,
\end{equation}
where the dependency on $m$ is omitted for simplicity.
The probability that exactly $w$ out of $N_w$ \glspl{sa-pull}  wake up follows a binomial distribution:
\begin{equation}
    P_{d}(w)=\binom{N_{w}}{w}P_{w}(\mb{V}_\mathrm{th})^{w}(1-P_{w}(\mb{V}_\mathrm{th}))^{N_{w}-w}.
    \label{eq:distribution_nwake}
\end{equation}
Similarly, the probability that exactly $u$ out of $N_{u}$ \glspl{sa-push} generate a packet in the $(m-1)$-th frame to transmit them on the shared portion of the $m$-th frame can be described as
\begin{equation}
    P_{u}(u) =\binom{N_{u}}{u} p_{\lambda}^u (1-p_{\lambda})^{N_{u}-u}.
    \label{eq:push_packet}
\end{equation}

\paragraph*{Push success probability}
Hereafter, we first derive the probability of successful access for \gls{sa-push}. Recall that each \gls{sa-push} having a packet to deliver starts transmitting at the beginning of shared slot $t_{c}(\alpha)$, contending the channel with active \glspl{sa-push} and with \glspl{sa-pull} who have not completed data transmission by $t_{c}(\alpha)$. 
Given $w$ the number of \glspl{sa-pull} woken up by a \gls{cowu} signaling, the probability that $r_w \le w$ \glspl{sa-pull} still have a packet to be transmitted after $\tau_w$ slots can be expressed as $P_{s}(w-r_{w}|w,  \tau_w)$ (see eq.~\eqref{eq:n_s_D}).
This implies that $r_{w} + u$ packets are contending the channel from slot $t_{c}(\alpha)$ to slot $L$. Hence, by applying again eq.~\eqref{eq:n_s_D}, the probability that $y$ nodes out of $r_{w} + u$ nodes successfully transmit within the duration of $\tau_u$ can be obtained by $P_{s}(y|r_{w}+u, \tau_u)$.

We are now interested in how many \glspl{sa-push} successfully transmit data among the $y$ nodes by the end of the $m$-th frame. 
The probability that $z_{u}$ \glspl{sa-push} ($0~\leq~z_{u}~\leq~y\leq~r_{w}+u$) out of $y$ nodes succeeds in data transmission, given $y$, $r_{w}$, $u$, can be derived by considering all possible combinations:
\begin{equation}
    P_{\beta}(z_{u}|y,u,r_{w})=\frac{\binom{u}{z_{u}}\binom{r_{w}}{y-z_{u}}}{\binom{r_{w}+u}{y}},
    \label{eq:beta_u}
\end{equation}
where the denominator represents all possible combinations for $y$ successful data transmission out of the $r_{w}+u$ available packets at the $t_c(\alpha)$ slot, while the numerator represents the number of combinations realizing $z_{u}$ \glspl{sa-push} successful data transmissions.
Then, we define the ratio of push successful data transmission given $u$ active \glspl{sa-push} at the $m$-th frame as 
\begin{equation}
    P_{\iota}(z_{u}|u)=
    \begin{cases}
        1, &\text{if}~u = 0,\\
        \frac{z_{u}}{u}, &\text{otherwise}.
    \end{cases}
    \label{eq:push_P_s_Z_U}
\end{equation}

Finally, using the law of total probability on eqs.~\eqref{eq:n_s_D}, \eqref{eq:distribution_nwake}, \eqref{eq:push_packet}, \eqref{eq:beta_u}, and~\eqref{eq:push_P_s_Z_U}, the push success probability at the $m$-th frame can be expressed by eq.~\eqref{eq:gamma_u_Push}.
% \begin{equation}\begin{split}
%     &\gamma_u(N_{w}, N_{u}, \mb{V}_{\mathrm{th}}, \lambda, \alpha)=\sum_{w=0}^{N_{w}}P_{d}(w)\sum_{u=0}^{N_{u}}P_{u}(u)\\&\sum_{r_{w}=0}^{w}P_{s}(w-r_{w}|w,t_{c}(\alpha))\sum_{y=1}^{u+r_{w}}P_{s}(y|u+r_{w},L-t_{c}(\alpha))\\&\sum_{z_{u}=0}^{\min(u,y)}P_{\beta}(z_{u}|y,u,r_{w})P_{\iota}(z_{u}|u).\label{eq:gamma_u_Push}
% \end{split}
% \end{equation}

\paragraph*{Accuracy of pull retrieved data}
To evaluate the retrieved accuracy defined in eq.~\eqref{eq:range_set}, we need to further derive the probability of $r_{w}$ \glspl{sa-pull} succeeding in their data transmission within the shared slots, assuming $r_{w} + u$ packets are available at slot $t_c(\alpha)$. Given $y$, $r_{w}$, and $u$, this can be derived as in~\eqref{eq:beta_u}:
\begin{equation}
    P_{\gamma}(r_{w}|y,u,r_{w}) = \frac{\binom{u}{y-r_{w}}}{\binom{r_{w}+u}{y}}.
    \label{eq:gamma_w}
\end{equation}
Applying the law of total probability on the eqs.~\eqref{eq:n_s_D}, \eqref{eq:distribution_nwake},~\eqref{eq:push_packet}, and~\eqref{eq:gamma_w}, the accuracy of range set for \gls{sa-pull} defined by eq.~\eqref{eq:range_set} can be computed as eq.~\eqref{eq:gamma_w_Pull}.
% \begin{equation}
% \begin{split}
% &\gamma_w(N_{w}, N_{u}, V_{\mathrm{th}}, \lambda, \alpha) =\sum_{w=0}^{N_{w}}P_{d}(w)\sum_{u=0}^{N_{u}}P_{u}(u)\\&\sum_{r_{w}=0}^{w}P_{s}(w-r_{w}|w,t_{c}(\alpha))\sum_{y=r_{w}}^{r_{w}+u}P_{s}(y|u+r_{w},L-t_{c}(\alpha))\\&P_{\gamma}(r_{w}|y,u,r_{w}).\label{eq:gamma_w_Pull}
% \end{split}
% \end{equation}

\paragraph*{Pull energy consumption}
Now, we will derive the total energy consumption for \glspl{sa-pull}, which is a function of the amount of time spent in the transmit and receive states. To this end, we first define the expected energy consumption for each state $s$ of the constructed Markov chain, denoted as $\bm{\Psi}= [\psi_{J}, \psi_{J-1}, \ldots, \psi_{0}]\T$, where $\psi_{s}$ is the expected amount of energy consumed when $s$ nodes are active and attempting to access the channel, i.e., there are $s$ packets left to transmit. Denoting the probability that $i$ out of $s$ packets are transmitted as $B(i)= \binom{s}{i}\, p^i\, (1-p)^{s-i}$ and its energy consumption as $e(i) = i T_{s} \xi_{T} + (s-i) T_{s} \xi_{R}$, the expected energy consumed for the state $s$, $\psi_{s}$, is
\begin{equation}
\ \psi_{s} =\sum_{i=0}^{s}e(i)B(i).\label{eq:Energy_cost_Fun}
\end{equation}
Finally, using the law of total probability on the eqs.~\eqref{eq:n_s_D}, \eqref{eq:distribution_nwake}, \eqref{eq:push_packet}, and~\eqref{eq:Energy_cost_Fun}, the total energy consumed by \glspl{sa-pull} during the $m$-th frame can be obtained from eq.~\eqref{eq:energy_tot}. In eq.~\eqref{eq:energy_tot}, the first term represents the total energy consumed by \glspl{sa-pull} during the reserved slots, while the second term is the pull total energy consumption during the shared slots. Furthermore, $c(r_{w}, u)$ represents the weight of the energy spent by the \glspl{sa-pull} during the shared slots, while contending the channel with \glspl{sa-push}. This is defined by dividing the number of \glspl{sa-pull} having an available packet at $t_c(\alpha)$ against the total available packets at $t_c(\alpha)$, i.e.,
\begin{equation}
\ c(r_{w}, u)=\begin{cases}\frac{r_{w}}{r_{w}+u},&\text{if}~r_{w} + u>0,\\0,&\text{otherwise}.\end{cases}
\end{equation}

\begin{figure*}[tbh]
    \centering
    \footnotesize
    \begin{equation}
        \gamma_u(N_{w}, N_{u}, \mb{V}_{\mathrm{th}}, \lambda, \alpha) = \sum_{w=0}^{N_{w}}P_{d}(w)\sum_{u=0}^{N_{u}}P_{u}(u)\sum_{r_{w}=0}^{w}P_{s}(w-r_{w}|w, \tau_w) \sum_{y=0}^{r_{w}+u}P_{s}(y|u+r_{w},\tau_u) \sum_{z_{u}=0}^{\min(u,y)}P_{\beta}(z_{u}|y,u,r_{w})P_{\iota}(z_{u}|u).
        \label{eq:gamma_u_Push}
    \end{equation}
    \begin{equation}
        \gamma_w(N_{w}, N_{u}, \mb{V}_{\mathrm{th}}, \lambda, \alpha) =\sum_{w=0}^{N_{w}}P_{d}(w)\sum_{u=0}^{N_{u}}P_{u}(u) \sum_{r_{w}=0}^{w}P_{s}(w-r_{w}|w,\tau_w) \sum_{y=r_{w}}^{r_{w}+u}P_{s}(y|u+r_{w},\tau_u) P_{\gamma}(r_{w}|y,u,r_{w}).
        \label{eq:gamma_w_Pull}
    \end{equation}
    \begin{equation}
    E_{\mathrm{tot}} (N_{w}, N_{u}, \mb{V}_{\mathrm{th}}, \lambda, \alpha) = \sum_{w = 0}^{N_{w}} {P_{d}(w)}\left[\sum_{t=1}^{\tau_w}\sum_{s=0}^{w}\phi_{s}(t)\psi_{s}+ \sum_{u = 0}^{N_{u}} P_{u}(u) \sum_{r_{w} =0}^{w} P_{s}(w-r_{w}|w,\tau_w) \, c(r_{w}, u)\sum_{t=1}^{\tau_u}\sum_{s=0}^{r_w+u}\phi_{s}(t)\psi_{s}\right].
    \label{eq:energy_tot}
    \end{equation}
    \par\noindent\rule{\textwidth}{0.4pt}
\end{figure*}
\normalsize

\section{Numerical Evaluation}
\label{sec:results}

This section evaluates the performance of the proposed pull/push coexistence scheme. 
In this evaluation, we generate a set of observed data for \gls{sa-pull} based on the distribution $g_{V}(v; m)$, and the generation of \gls{sa-push} packets based on the \gls{ppp}. The procedure of wake-up control, data collection and data transmission following $p$-persistent \gls{csma} operations are simulated, according to Sec.~\ref{sec:system_model}.
For simplicity, we assume that the observed process follows a uniform distribution, i.e., $g_{V}(v; m) = \frac{1}{V_{\mathrm{max}}-V_{\mathrm{min}}}$, for $v\in[V_\mathrm{min}, V_\mathrm{max}]$, $0$ otherwise.
The values of common parameters are shown in Table~\ref{table:para}. 

\begin{table}[bt]
\caption{Common parameters for simulations.}
    \footnotesize
    \centering    
    \begin{tabular}{c|c}
        \toprule
        Parameters & Values  \\
        \midrule
        Data transmission rate & 100~kbps \\ 
        Time slot duration $T_s$ & $3.2$ [ms]~\cite{TGCN_Content}\\
        Power consumption while transmitting~$\xi_{T}$ & 55 [mW]~\cite{tamura2019low}\\ 
        Power consumption while receiving~$\xi_{R}$ & 50~[mW]~\cite{tamura2019low}\\
        Distribution of observed value~$[V_{\mathrm{min}},V_{\mathrm{max}}]$ & $[0,1]$ \\ 
        Transmission probability $p$    & 0.0606~\cite{TGCN_Content} \\ 
        \bottomrule
    \end{tabular}
    \label{table:para}    
    \vspace{-4mm}
\end{table}
\normalsize
As a baseline scheme, we apply a \gls{rr} as in \cite{shiraishi2022query}, in which each \gls{sa-pull} transmits its observed data in the first $N_w$ slots in a predetermined order, while the \gls{sa-push} transmits data using $p$-persistent \gls{csma} in the subsequent $L - N_{w}$ slots. In \gls{rr}, as the \gls{bs} can collect data from all \glspl{sa-pull} without collisions, the pull retrieved data accuracy is always 1, while the total energy consumption is
\begin{equation}
    E_{\mathrm{tot}}^{\mathrm{RR}}(N_{w}) = N_{w}T_{s}\xi_{T}.\label{eq:ene_RR}
\end{equation}
On the other hand, the push success probability results in
\begin{equation}
    \gamma_{u}^\mathrm{RR} = \sum_{u=0}^{N_{u}}P_{u}(u)\sum_{z_{u}=0}^{u}P_{s}(z_{u}|u,L-N_{w})P_{\iota}(z_{u}|u).
\end{equation}

\subsection{Trade-off between pull and push performance}\label{sec:Trade_off}

 \begin{figure*}[htb]
     \centering
     \begin{tikzpicture}

\definecolor{crimson2143940}{RGB}{214,39,40}
\definecolor{darkgray176}{RGB}{176,176,176}
\definecolor{darkorange25512714}{RGB}{255,127,14}
\definecolor{forestgreen4416044}{RGB}{44,160,44}
\definecolor{lightgray204}{RGB}{204,204,204}
\definecolor{mediumpurple148103189}{RGB}{148,103,189}

\def\hsep{1.2cm}
\def\vsep{1cm}
\def\vside{2.8cm}
\def\hside{0.25\textwidth}

\begin{groupplot}[
group style={group name=system, group size=3 by 1,  horizontal sep=\hsep, vertical sep=\vsep, xlabels at=edge bottom}, 
title style={at={(0.5,0.85)}},
anchor=south east,
height=\vside,
width=\hside,
scale only axis,
legend style={  
  at={(0.5, 1.05)}, 
  draw=none,
  fill opacity=0,
  anchor=south,  
  /tikz/every even column/.append style={column sep=0.3cm}
},
legend columns=-1,
xmajorgrids,
xlabel={$\alpha$},
xmin=0, xmax=1,
xtick={0,0.5,1},
xticklabels={
  \(\displaystyle {0.0}\),
  \(\displaystyle {0.5}\),
  \(\displaystyle {1.0}\),  
},
ymin=0, ymax=1,
ytick={0,0.25,0.5,0.75,1,1.5,2.},
ymajorgrids,
yticklabels={
  \(\displaystyle {0.0}\),
  \(\displaystyle {0.25}\),
  \(\displaystyle {0.50}\),
  \(\displaystyle {0.75}\),
  \(\displaystyle {1.0}\),
  \(\displaystyle {1.5}\),
  \(\displaystyle {2.0}\),  
},
ylabel shift=-4pt,
xlabel shift=-3pt,
]
\nextgroupplot[ylabel={$\gamma_w$}]
\addplot [thick, orange, dashdotted, forget plot]
table {%
0 0.0244868834947786
0.2 0.032679271749433
0.4 0.0439423030237447
0.6 0.0596586844634698
0.8 0.0815725357424052
1 0.111582498753892
};

\addplot [thick, orange, mark=asterisk, mark size=1.5, mark options={solid, fill=white}, only marks, forget plot]
table {%
0 0.02458
0.2 0.0321
0.4 0.04262
0.6 0.06078
0.8 0.0806
1 0.10996
};

\addplot [thick, cyan, forget plot]
table {%
0 0.131461353436132
0.2 0.188867650462157
0.4 0.260769229361492
0.6 0.353679022405222
0.8 0.467655601225398
1 0.591556390534104
};

\addplot [thick, cyan, mark=*, mark size=1.5, mark options={solid, fill=white}, only marks, forget plot]
table {%
0 0.1341
0.2 0.18708
0.4 0.26038
0.6 0.35462
0.8 0.46726
1 0.59376
};

\addplot [thick, red, dashed, forget plot]
table {%
0 0.364144503559581
0.2 0.480065819070288
0.4 0.583591116712246
0.6 0.693496884612498
0.8 0.802817829768053
1 0.889450380459957
};

\addplot [thick, red, mark=square*, mark size=1.5, mark options={solid, fill=white}, only marks, forget plot]
table {%
0 0.36
0.2 0.48138
0.4 0.57926
0.6 0.69028
0.8 0.80584
1 0.88994
};

\nextgroupplot[ylabel={$\gamma_u$}]
\addplot [thick, orange, dashdotted]
table {%
0 0.482492822070495
0.2 0.419041447764938
0.4 0.339867235609827
0.6 0.243676742876929
0.8 0.130110326505547
1 1.63737713059082e-07
};
\addlegendentry{Theory $L = 25$}
\addplot [thick, orange, mark=asterisk, mark size=1.5, mark options={solid, fill=white}, only marks]
table {%
0 0.483586396940784
0.2 0.417293549180123
0.4 0.340055565749553
0.6 0.244505657961899
0.8 0.129911262492817
1 0
};
\addlegendentry{Simulation $L = 25$}
\addplot [thick, cyan]
table {%
0 0.692986576136105
0.2 0.631237780778263
0.4 0.528742871986483
0.6 0.383149899557559
0.8 0.201836273015811
1 2.68100386778181e-14
};
\addlegendentry{Theory $L = 50$}
\addplot [thick, cyan, mark=*, mark size=1.5, mark options={solid, fill=white}, only marks]
table {%
0 0.692291691093657
0.2 0.631266373933591
0.4 0.529006333854986
0.6 0.382748482508731
0.8 0.201789279725889
1 0
};
\addlegendentry{Simulation $L = 50$}
\addplot [thick, red, dashed]
table {%
0 0.846370319522698
0.2 0.801588873765624
0.4 0.691990723148563
0.6 0.505246924312328
0.8 0.260783564397036
1 4.38981442013142e-21
};
\addlegendentry{Theory $L = 75$}
\addplot [thick, red, mark=square*, mark size=1.5, mark options={solid, fill=white}, only marks]
table {%
0 0.845949807743596
0.2 0.80184674758284
0.4 0.691384376817291
0.6 0.505398964844298
0.8 0.260799911059918
1 0
};
\addlegendentry{Simulation $L = 75$}

\nextgroupplot[
ylabel={$E_{\mathrm{tot}}$ [J]},
ymin=0.017, ymax=0.06,
ytick={0.02,0.03, 0.04,0.05,0.06,0.07},
yticklabels={
  \(\displaystyle {2}\),
  \(\displaystyle {3}\),
  \(\displaystyle {4}\),
  \(\displaystyle {5}\),
  \(\displaystyle {6}\),
  \(\displaystyle {7}\),  
},
]
\addplot [thick, orange, dashdotted, forget plot]
table {%
0 0.0232827642654975
0.2 0.021404985175303
0.4 0.0200814099672573
0.6 0.0192358301020755
0.8 0.0187925726352417
1 0.0186781620414106
};
% \addlegendentry{\scriptsize$\text{Theory $L$ = 25}$}
\addplot [thick, orange, mark=asterisk, mark size=1.5, mark options={solid}, only marks, forget plot]
table {%
0 0.02323625056
0.2 0.02148739072
0.4 0.0200429875200001
0.6 0.0192146908800001
0.8 0.0187811884800002
1 0.0187670656000001
};
% \addlegendentry{\scriptsize$\text{Simulation $L$ = 25}$}
\addplot [thick, cyan, forget plot]
table {%
0 0.0400647562897423
0.2 0.0320630663099675
0.4 0.0274183958139315
0.6 0.0250324095923142
0.8 0.024038959831378
1 0.023820429045878
};
% \addlegendentry{\scriptsize$\text{Theory $L$ = 50}$}
\addplot [thick, cyan, mark=*, mark size=1.5, mark options={solid, fill=white}, only marks, forget plot]
table {%
0 0.0400161977600002
0.2 0.0320392022400001
0.4 0.0274008518399999
0.6 0.0250109891200001
0.8 0.02408010432
1 0.0237866230399997
};
% \addlegendentry{\scriptsize$\text{Simulation $L$ = 50}$}
\addplot [thick, red, dashed, forget plot]
table {%
0 0.0517134436036729
0.2 0.035985725807388
0.4 0.028847320409736
0.6 0.0260405575110342
0.8 0.0251495812235178
1 0.0249953976273937
};
% \addlegendentry{\scriptsize$\text{Theory $L$ = 75}$}
\addplot [thick, red, mark=square*, mark size=1.5, mark options={solid, fill=white}, only marks, forget plot]
table {%
0 0.0518312284800002
0.2 0.0360008400000002
0.4 0.0289064553600001
0.6 0.0260643363200001
0.8 0.0251005798399999
1 0.0249737292800001
};
% \addlegendentry{\scriptsize$\text{Simulation $L$ = 75}$}

\end{groupplot}

%% Subcaptions
\node[text width=\hside, align=center, anchor=north] at ([yshift=-0.6cm]system c1r1.south) {\protect\subcaption{$\gamma_w$ vs. $\alpha$ for \gls{sa-pull}.\protect\label{fig:top_k_acc_against_alpha}}};

\node[text width=\hside, align=center, anchor=north] at ([yshift=-0.6cm]system c2r1.south) {\protect\subcaption{$\gamma_u$ vs. $\alpha$ for \gls{sa-push}.\protect\label{fig:P_s_alpha}}};

\node[text width=\hside, align=center, anchor=north] at ([yshift=-0.6cm]system c3r1.south) {\protect\subcaption{$E_\mathrm{tot}$ vs. $\alpha$ for \gls{sa-pull}. \protect\label{fig:ene_alpha}}};

\end{tikzpicture}
     \caption{The system level performance for the \glspl{sa-pull} and \glspl{sa-push} vs. $\alpha$, applying \gls{cowu}.}
     \label{fig:system-performance}
     \vspace{-4mm}
 \end{figure*}
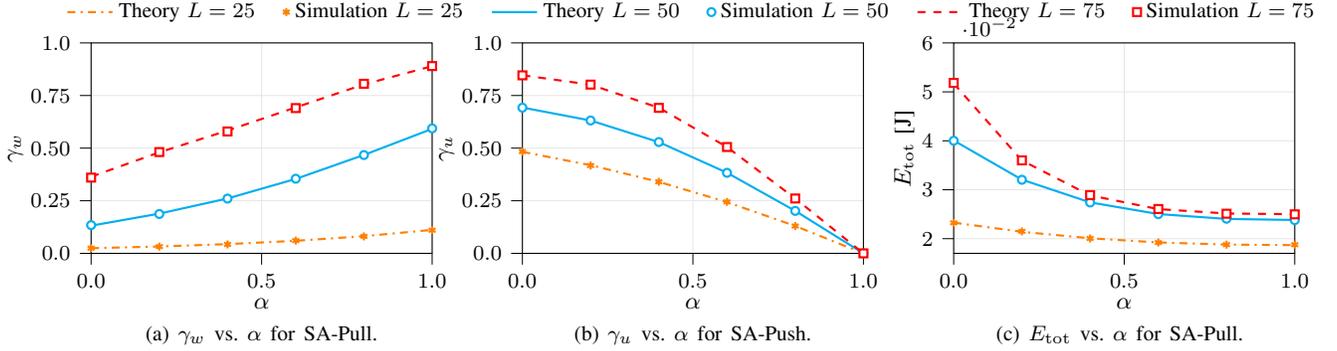
 
Figs.~\ref{fig:top_k_acc_against_alpha},~\ref{fig:P_s_alpha}, and \ref{fig:ene_alpha} show the pull retrieval data accuracy, the push success probability, and the pull energy consumption as a function of $\alpha$, respectively, showing the results obtained by theoretical analysis and computer simulation, with parameter $N_{w} = 25$, $N_{u} = 25$, $\mb{V}_\mathrm{th} = [0.6, 0.9]$, and $\lambda = 0.025$. First, we can see that the results obtained by our theoretical analysis coincide with those obtained by simulation, validating our analysis. Next, from Fig.~\ref{fig:top_k_acc_against_alpha}, we can see that $\gamma_w$ increases as the value of $\alpha$ becomes larger. When $\alpha$ is small, the \glspl{sa-push} start data transmission earlier in the frame, and the remaining \glspl{sa-pull} need to contend the channel for a longer time, increasing the chances of inter-class collisions. 
On the other hand, from Fig.~\ref{fig:P_s_alpha}, $\gamma_u$ becomes smaller as $\alpha$ increases because of fewer opportunities for transmission. 
 Finally, from Fig.~\ref{fig:ene_alpha}, we can see that $E_{\mathrm{tot}}$ becomes smaller as the value of $\alpha$ becomes larger. This is because as the $\alpha$ increases, \glspl{sa-pull} are more likely to successfully transmit -- and thus turn-off -- at the earlier timing thanks to the smaller duration of shared slots, mitigating the amount of collisions. 
These results clearly illustrate the basic trade-off between the pull and push performance generated by the choice of $\alpha$. %, and the importance of \gls{bs} allocating resources considering both requirements. 

\subsection{Acceptable arrival rate $\lambda$}
\label{sec:arrival_rate}

To compare the performance of \gls{cowu} and \gls{rr} as fairly as possible, we set a threshold $\gamma_{\mathrm{th}}$ as the minimum value of both push success probability and pull retrieval accuracy. Under these constraints, we compute the maximum acceptable $\lambda$ that can be supported by the system, namely $\lambda_\mathrm{max}$, jointly with the set of $\{\alpha\}$ that makes the problem feasible. That is
\begin{align}
    \lambda_\text{max} = & \max_{\{\alpha, \lambda\}} \lambda \label{eq:max_lambda}\\
    \text{s.t.} &\quad \gamma_{w}(N_{w}, N_{u}, \mb{V}_\mathrm{th}, \lambda, \alpha) \geq \gamma_\mathrm{th},  \tag{\ref{eq:max_lambda}.a} \label{eq:pull_cond} \\
    & \quad \gamma_{u}(N_{w}, N_{u}, \mb{V}_\mathrm{th}, \lambda, \alpha) \geq \gamma_\mathrm{th}. \tag{\ref{eq:max_lambda}.b} \label{eq:push_cond} 
\end{align}
Given $N_w$, $N_u$, and $\mb{V}_\mathrm{th}$, problem~\eqref{eq:max_lambda} is solved employing a grid-search varying the value of $\alpha$ within a range of [0, 1] and a step of 0.05, and value of $\lambda$ within a range of [0.005, 0.050] and a step of 0.005.

\begin{figure}[t!]
    \centering
    % This file was created with tikzplotlib v0.10.1.
\begin{tikzpicture}

\definecolor{crimson2143940}{RGB}{214,39,40}
\definecolor{darkgray176}{RGB}{176,176,176}
\definecolor{darkorange25512714}{RGB}{255,127,14}
\definecolor{forestgreen4416044}{RGB}{44,160,44}
\definecolor{mediumpurple148103189}{RGB}{148,103,189}
\definecolor{sienna1408675}{RGB}{140,86,75}
\definecolor{steelblue31119180}{RGB}{31,119,180}

\begin{axis}[
width=0.50\columnwidth,
height=2.5cm,
scale only axis,
xmajorgrids,
ymajorgrids,
tick align=outside,
tick pos=left,
xlabel={$N_{w}$},
xmin=3, xmax=47,
ylabel={$\lambda_{\mathrm{max}}$ [packet/slot]},
ymin=0, ymax=0.055,
legend style={at={(1,0.5)},
    anchor=west, 
    legend columns=1,
    draw=none,
    fill opacity=0,
    /tikz/every even column/.append style={column sep=0.3cm}},
ylabel shift = -3pt,
xlabel shift = -3pt,
]
\addplot [thick, red, mark=square*, mark size=1.5, mark options={solid, fill=white}]
table {%
5 0.05
15 0.05
25 0.05
35 0.05
45 0.04
};
\addlegendentry{CoWu ($N_{u} = 15$)}

\addplot [thick,black,  mark=diamond*, mark size=1.5, mark options={solid, fill=white}]
table {%
5 0.05
15 0.02
25 0
35 0
45 0
};
\addlegendentry{RR ($N_{u} = 15$)}

\addplot [thick,red, dashed, mark=star, mark size=1.5,mark options={solid, fill=white}]
table {%
5 0.03
15 0.03
25 0.025
35 0.02
45 0.01
};
\addlegendentry{CoWu ($N_{u} = 25$)}

\addplot [thick,black, dashed, mark=triangle*, mark size=1.5,mark options={solid, fill=white}]
table {%
5 0.02
15 0.01
25 0
35 0
45 0
};
\addlegendentry{RR ($N_{u} = 25$)}

\addplot [thick,red, dotted, mark=*, mark size=1.5, mark options={solid, fill=white}]
table {%
5 0.015
15 0.015
25 0.015
35 0.01
45 0.005
};
\addlegendentry{CoWu ($N_{u} = 35$)}

\addplot [thick,black,dotted,  mark=+, mark size=1.5,,mark options={solid, fill=white}]
table {%
5 0.01
15 0.005
25 0
35 0
45 0
};
\addlegendentry{RR ($N_{u} = 35$)}
\end{axis}

\end{tikzpicture}
    \caption{Acceptable traffic $\lambda_\mathrm{max}$ vs. $N_{w}$.}
    \label{fig:Region_lambda}
    \vspace{-4mm}
\end{figure}
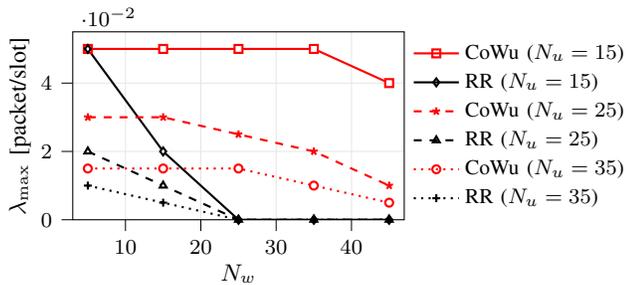

Fig.~\ref{fig:Region_lambda} shows $\lambda_\mathrm{max}$ as a function of the number of \glspl{sa-pull} $N_{w}$ for \gls{cowu} and \gls{rr}, where we set $N_{u} \in\{ 15, 25, 35\}$, $L = 50$, $\mb{V}_\mathrm{th} = [0.94, 0.98]$, and $\gamma_{\mathrm{th}} = 0.8$. The acceptable traffic drops for both schemes as $N_{w}$ increases because the available time resources for \glspl{sa-push} have to decrease to support the reliable data transmission of a higher number of \glspl{sa-pull}, i.e., $\alpha$ increases to satisfy constraint~\eqref{eq:pull_cond}. Moreover, for both schemes, $\lambda_\mathrm{max}$ decreases as $N_{u}$ increases due to the higher probability of collisions.
The degradation of $\lambda_\mathrm{max}$ is much faster for \gls{rr}, due to the fact that the available resources for \glspl{sa-push} become significantly smaller as $N_{w}$ becomes larger. Conversely, in \gls{cowu}, only a subset of \glspl{sa-pull} observing values within $\mb{V}_\mathrm{th}$ wake up and transmit data. This enables the \gls{bs} to set a small value of $\alpha$, increasing $\lambda_\mathrm{max}$. Finally, \gls{cowu} achieves a better acceptable traffic metric than \gls{rr} for any given value of $N_{w}$ and $N_u$.

\subsection{Pull energy consumption}

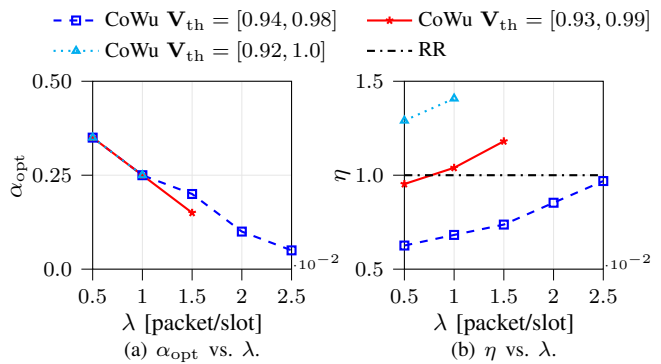
\begin{figure}[t!]
    \centering
    \begin{tikzpicture}

\definecolor{crimson2143940}{RGB}{214,39,40}
\definecolor{darkgray176}{RGB}{176,176,176}
\definecolor{darkorange25512714}{RGB}{255,127,14}
\definecolor{forestgreen4416044}{RGB}{44,160,44}
\definecolor{lightgray204}{RGB}{204,204,204}
\definecolor{mediumpurple148103189}{RGB}{148,103,189}

\def\hsep{1.5cm}
\def\vsep{1cm}
\def\vside{2.5cm}
\def\hside{0.3\columnwidth}

\begin{groupplot}[
group style={group name=syssys, group size=2 by 1,  horizontal sep=\hsep, vertical sep=\vsep,  xlabels at=edge bottom}, 
anchor=south east,
height=\vside,
width=\hside,
scale only axis,
legend style={  
  at={(-0.25, 1.05)}, 
  legend columns=2,
  draw=none,
  fill opacity=0,
  anchor=south,  
  /tikz/every even column/.append style={column sep=0.3cm}
},
xmajorgrids,
xlabel={$\lambda$ [packet/slot]},
xmin=0.005, xmax=0.025,
ymajorgrids,
ylabel shift=-5pt,
xlabel shift=-3pt,
every x tick scale label/.style={at={(rel axis cs:1,0)},anchor=south west,inner sep=0pt, font=\tiny},
]

\nextgroupplot[
ylabel={$\alpha_{\mathrm{opt}}$},
ymin=0, ymax=0.5,
ytick={0,0.25,0.5},
yticklabels={
  \(\displaystyle {0.0}\),
  \(\displaystyle {0.25}\),
  \(\displaystyle {0.50}\),  
},]
\addplot [thick, blue,dashed, mark=square,  mark size=1.5, mark options={solid, fill=white}]
table {%
0.005 0.35
0.01 0.25
0.015 0.2
0.02 0.1
0.025 0.05
};

\addplot [thick, red, mark=star, mark size=1.5, mark options={solid, fill=white}]
table {%
0.005 0.35
0.01 0.25
0.015 0.15
};

\addplot [thick, cyan,dotted, mark=triangle, mark size=1.5, mark options={solid, fill=white}]
table {%
0.005 0.35
0.01 0.25
};

\nextgroupplot[
ylabel={$\eta$},
ymin=0.5, ymax=1.5,
ytick={0.5, 1, 1.5},
yticklabels={
  \(\displaystyle {0.5}\),
  \(\displaystyle {1.0}\),
  \(\displaystyle {1.5}\),  
},
]

\addplot [thick, blue,dashed, mark=square,  mark size=1.5, mark options={solid, fill=white}]
table {%
0.005 0.625749132016461
0.01 0.68216232227729
0.015 0.737240822687484
0.02 0.853491823509066
0.025 0.96893011820783
};
\addlegendentry{CoWu~$\mathbf{V}_\mathrm{th} = [0.94, 0.98]$}

\addplot [thick, red, mark=star, mark size=1.5, mark options={solid, fill=white}]
table {%
0.005 0.953103877380311
0.01 1.03950544530316
0.015 1.18068538653918
%0.02 nan
%0.025 nan
%0.03 nan
};
\addlegendentry{CoWu~$\mathbf{V}_\mathrm{th} = [0.93, 0.99]$}

\addplot [thick, cyan,dotted, mark=triangle, mark size=1.5, mark options={solid, fill=white}]
table {%
0.005 1.29045285511167
0.01 1.40805919935895
%0.015 nan
%0.02 nan
%0.025 nan
%0.03 nan
};
\addlegendentry{CoWu~$\mathbf{V}_\mathrm{th} = [0.92, 1.0]$}

\addplot [thick, black, dash dot]
table {%
0.005 1
0.030 1
};
\addlegendentry{RR}

% Draw a vertical dashed line
%\draw[thick, black, dashed] (axis cs:0.01,0) -- (axis cs:0.01,1.5);

% Draw a vertical dashed line with an arrow at the end
%\draw[thick, black,  -stealth] (axis cs:0.015,0.54)--(axis cs:0.01,0.55);

% Add a label
%\node[align=center, font=\small, black] at (axis cs:0.018,0.58) {$\lambda_{\mathrm{max}}^{\mathrm{RR}}$};

\end{groupplot}
%% Subcaptions
\node[text width=\hside, align=center, anchor=north] at ([yshift=-0.6cm]syssys c1r1.south) {\protect\subcaption{$\alpha_{\mathrm{opt}}$ vs. $\lambda$.\protect\label{fig:opt_alpha}}};

\node[text width=\hside, align=center, anchor=north] at ([yshift=-0.6cm]syssys c2r1.south) {\protect\subcaption{$\eta$ vs. $\lambda$.\protect\label{fig:eta}}};

\end{tikzpicture}
     \vspace{-4mm}
    \caption{$\alpha_{\mathrm{opt}}$ and $\eta$ vs. $\lambda$  [packets/slots], applying \gls{cowu}.}
    \label{fig:Region_ene}
    \vspace{-4mm}
\end{figure}

%Here, we compare the energy consumption of \gls{cowu} and \gls{rr}, defining the energy consumption ratio as
%\begin{equation}
%    \eta=\frac{E_{\mathrm{tot}}(N_{w}, N_{u}, \mb{V}_{\mathrm{th}}, \lambda_\mathrm{max}, \alpha_{\mathrm{opt}}(\lambda_{\mathrm{max}}))}{E_{\mathrm{tot}}^{\mathrm{RR}}(N_{w})},\label{eq:eta}
%\end{equation}
%\begin{equation}
%    \eta=\frac{E_{\mathrm{tot}}(N_{w}, N_{u}, \mb{V}_{\mathrm{th}}, \lambda_\mathrm{max}, \alpha_{\mathrm{opt}}(\lambda_{\mathrm{max}}))}{E_{\mathrm{tot}}^{\mathrm{RR}}(N_{w})},\label{eq:eta}
%\end{equation}
%where $\alpha_\mathrm{opt}$ is the value of $\alpha$ minimizing the energy consumption under the same conditions of Sec.~\ref{sec:arrival_rate}, i.e.,
As the pull energy consumption depends on the value of $\alpha$ (c.f. Sec.~\ref{sec:Trade_off}), we first obtain the optimal value of $\alpha$ minimizing it under the same conditions of Sec.~\ref{sec:arrival_rate} for different values of $\lambda$, i.e.,
\begin{equation}
    %\begin{aligned}
     \alpha_{\mathrm{opt}} (\lambda) = \argmin_{\alpha} E_{\mathrm{tot}}(\mb{V}_{\mathrm{th}}, \lambda, \alpha) ~~~\text{s.t.}~\eqref{eq:pull_cond},~\eqref{eq:push_cond}.\label{eq:Eopt}
    %\end{aligned}     
\end{equation}
Fig.~\ref{fig:opt_alpha} shows $\alpha_{\mathrm{opt}}$ values against $\lambda$, where we set $N_{w} = 25$, $N_{u} = 25$, $L = 50$, $\gamma_{\mathrm{th}} = 0.8$, and $\mb{V}_{\mathrm{th}} \in \{[0.94, 0.98],\,[0.93, 0.99],\,[0.92, 1.0]\}$. The value of $\alpha_{\mathrm{opt}}$ becomes smaller as $\lambda$ increases, because the \gls{bs} needs to increase $\tau_{u}$ to satisfy the condition of eq.~\eqref{eq:push_cond}. Next, Fig.~\ref{fig:eta} shows the energy consumption ratio between \gls{cowu} and \gls{rr} approaches, i.e., $\eta = E_{\mathrm{tot}}(\mb{V}_{\mathrm{th}}, \lambda, \alpha_{\mathrm{opt}}(\lambda)) / E_{\mathrm{tot}}^{\mathrm{RR}}$, against push-based traffic $\lambda$. 
$\eta$ becomes lower than 1 only if the energy consumption of \glspl{sa-pull} for \gls{cowu} is smaller than the \gls{rr} one, indicating performance gain in terms of energy efficiency. In Fig.~\ref{fig:eta}, the values of $\eta$ are not plotted, if the system can not support requirements~\eqref{eq:pull_cond}, and~\eqref{eq:push_cond}. We observe the value of $\eta$ increases as $\lambda$ becomes larger, because the \glspl{sa-pull} experiences more inter-class collision due to the shorter $\tau_{w}$, that eventually adds more number of available \glspl{sa-push} packets. Furthermore, $\eta$ becomes larger as the range of requested content becomes larger. This is simply because more \glspl{sa-pull} wakes up against a \gls{cowu} signal, which increases the time to complete data transmission due to the contentions. Nevertheless, we can see that \gls{cowu} can outperform \gls{rr} (up to 38 $\%$ in terms of pull energy consumption) for specific query intervals, e.g., $\mb{V}_{\mathrm{th}} = [0.94, 0.98]$, while maintaining high accuracy of pull retrieved data and high push success probability. This proves the effectiveness of the use of \gls{cowu} to ask for relevant data depending on the system's traffic.

\section{Conclusions}
We have investigated a pull/push coexistence scenario, where \gls{iot} devices equipped with wake-up receivers respond to \gls{cowu} signal sent by the \gls{bs} for a high data retrieval accuracy. We established theoretical models and derived optimal operation points for an energy-efficient transmission strategy between \gls{sa-pull} and \gls{sa-push} that ensures a high transmission success rate for \gls{sa-push}. Our numerical results have shown the basic trade-off between the accuracy of push success probability and pull retrieved data accuracy, and energy consumption. % Future work includes considerations on the timing of wake-up control for the pull/push coexistence scenario.

% In this paper, focusing on \gls{iot} data collection from the pull and push coexistence system, we have investigated how we can retrieve the range-set from the \gls{iot} devices equipped with the wake-up receiver with high energy-efficiency and high accuracy while maintaining high transmission success rate for the nodes serving push-based communications. In order to allocate the limited wireless resources (spectrum and energy) only for nodes having valuable data, we have applied \gls{cowu} and have investigated its performance with theoretical analysis. Our numerical results have characterized the basic trade-off between the accuracy of retrieved range-set for \gls{sa-pull} and transmission success rate for \gls{sa-push} through the amount of allocated time resources. Further, we have revealed the region where the applying \gls{cowu} for the coexistence scenario outperforms round robin scheme in terms of total energy consumption and required target metric for both push and pull devices. Future work includes the investigation of the timing of wake-up control for the pull and push coexistence scenario. 

%%% BIBLIOGRAPHY
\bibliographystyle{IEEEtranNoURL}
\bibliography{IEEEabrv,Ref_2}

\begin{thebibliography}{10}
\providecommand{\url}[1]{#1}
\csname url@samestyle\endcsname
\providecommand{\newblock}{\relax}
\providecommand{\bibinfo}[2]{#2}
\providecommand{\BIBentrySTDinterwordspacing}{\spaceskip=0pt\relax}
\providecommand{\BIBentryALTinterwordstretchfactor}{4}
\providecommand{\BIBentryALTinterwordspacing}{\spaceskip=\fontdimen2\font plus
\BIBentryALTinterwordstretchfactor\fontdimen3\font minus
  \fontdimen4\font\relax}
\providecommand{\BIBforeignlanguage}[2]{{%
\expandafter\ifx\csname l@#1\endcsname\relax
\typeout{** WARNING: IEEEtran.bst: No hyphenation pattern has been}%
\typeout{** loaded for the language `#1'. Using the pattern for}%
\typeout{** the default language instead.}%
\else
\language=\csname l@#1\endcsname
\fi
#2}}
\providecommand{\BIBdecl}{\relax}
\BIBdecl

\bibitem{Chettri2020:iot-survey}
L.~Chettri and R.~Bera, ``A comprehensive survey on internet of things ({IoT})
  toward {5G} wireless systems,'' \emph{IEEE Internet Things J.}, vol.~7,
  no.~1, pp. 16--32, 2020.

\bibitem{Goldreich2012:goal-theory}
\BIBentryALTinterwordspacing
O.~Goldreich \emph{et~al.}, ``A theory of goal-oriented communication,''
  \emph{J. ACM}, vol.~59, no.~2, may 2012.
\BIBentrySTDinterwordspacing

\bibitem{gunduz2022beyond}
D.~G{\"u}nd{\"u}z \emph{et~al.}, ``Beyond transmitting bits: Context,
  semantics, and task-oriented communications,'' \emph{IEEE J. Sel. Areas
  Commun.}, vol.~41, no.~1, pp. 5--41, 2022.

\bibitem{Chiariotti2022:qAoI}
F.~Chiariotti \emph{et~al.}, ``Query age of information: Freshness in
  pull-based communication,'' \emph{IEEE Trans. Commun.}, vol.~70, no.~3, pp.
  1606--1622, 2022.

\bibitem{shiraishi2022query}
J.~Shiraishi \emph{et~al.}, ``Query timing analysis for content-based wake-up
  realizing informative {IoT} data collection,'' \emph{IEEE Wireless Commun.
  Lett.}, vol.~12, no.~2, pp. 327--331, 2023.

\bibitem{piyare2017ultra}
R.~Piyare \emph{et~al.}, ``Ultra low power wake-up radios: A hardware and
  networking survey,'' \emph{IEEE Commun. Surv. Tut.}, vol.~19, no.~4, pp.
  2117--2157, 2017.

\bibitem{IEICE-yomo}
H.~Yomo \emph{et~al.}, ``{ROD-SAN}: Energy-efficient and high-response wireless
  sensor and actuator networks employing wake-up receiver,'' \emph{IEICE Trans.
  Commun.}, vol. E99-B, no.~9, pp. 1998--2008, Sept. 2016.

\bibitem{3GPP_standard}
3GPP, ``Study on low-power wake up signal and receiver for {NR} (release 18),''
  3GPP, Tech. Rep. 38.869, V2.0.0, 2023.

\bibitem{wagner2023low}
S.~Wagner \emph{et~al.}, ``Low-power wake-up signal design in {3GPP} release
  18,'' in \emph{2023 IEEE Conf. Standards Commun. Netw. (CSCN)}.\hskip 1em
  plus 0.5em minus 0.4em\relax IEEE, 2023, pp. 222--227.

\bibitem{hoglund20243gpp}
A.~Hoglund \emph{et~al.}, ``{3GPP} release 18 wake-up receiver: Feature
  overview and evaluations,'' \emph{arXiv preprint arXiv:2401.03333}, 2024.

\bibitem{TGCN_Content}
J.~Shiraishi \emph{et~al.}, ``Content-based wake-up for top-k query in wireless
  sensor networks,'' \emph{IEEE Trans. Green Commun. Netw.}, vol.~5, no.~1, pp.
  362--377, 2021.

\bibitem{Sara_Pull_and_push}
S.~Cavallero \emph{et~al.}, ``Coexistence of pull and push communication in
  wireless access for {IoT} devices,'' \emph{arXiv preprint arXiv:2404.07650},
  2024.

\bibitem{talli2024push}
P.~Talli \emph{et~al.}, ``Push- and pull-based effective communication in
  cyber-physical systems,'' \emph{arXiv preprint arXiv:2401.10921}, 2024.

\bibitem{tamura2019low}
N.~Tamura \emph{et~al.}, ``Low-overhead wake-up control for wireless sensor
  networks employing wake-up receivers,'' \emph{IEICE Trans. Commun.}, vol.
  102, no.~4, pp. 732--740, 2019.

\end{thebibliography}

\end{document}